# BaTh$_2$Fe$_4$As$_4$(N$_{0.7}$O$_{0.3}$)$_2$：An Iron-Based Superconductor Stabilized by Inter-Block-Layer Charge Transfer

Ye-Ting Shao[1], Zhi-Cheng Wang[1], Bai-Zhuo Li[1], Si-Qi Wu[1], Ji-Feng Wu[2], Zhi Ren[2], Su-Wen Qiu[3], Can Rao[3], Cao Wang[4] and Guang-Han Cao[1,5*]

[1] Department of Physics, Zhejiang Province Key Laboratory of Quantum Technology and Device, and Stat Key Lab of Silicon Materials, Zhejiang University, Hangzhou 310027, China

[2] School of Sciences, Westlake Institute for Advanced Study, Westlake University, Hangzhou 310064, China

[3] School of Earth Sciences, Zhejiang University, Hangzhou 310027, China

[4] Department of Physics, Shandong University of Technology, Zibo 255049, China

[5] Collaborative Innovation Centre of Advanced Microstructures, Nanjing University, Nanjing 210093, China

* Corresponding author (email: ghcao@zju.edu.cn)

**Abstract**

Recently, An electron-doped 12442-type iron-based superconductor BaTh$_2$Fe$_4$As$_4$(N$_{0.7}$O$_{0.3}$)$_2$ has been successfully synthesized with high-temperature solid-state reactions on basis of a structural design. The inter-block-layer charge transfer between the constituent units of "BaFe$_2$As$_2$" and "ThFeAsN$_{0.7}$O$_{0.3}$" was found to be essential to stabilize the target compound. Dominant electron-type conduction and bulk superconducting transition at ∼22 K were demonstrated.

**Introduction.** The discovery of superconductivity at $T_c$ = 26 K in fluorine-doped LaFeAsO[1] has led to the subsequent findings of dozens of new superconducting materials in the class of Fe-based superconductors (FeSCs).[2,3] All those FeSCs were found to possess antifluorite-like $Fe_2X_2$ ($X$ = As, Se) layers which are believed to be the crucial structural motif for the emergence of high-temperature superconductivity. This common feature makes it possible to design a new FeSC with the crystal chemistry.[3-5] For example, with the consideration of the intergrowth of $BaFe_2As_2$ and $BaTi_2As_2O$, the new FeSC $Ba_2Fe_2Ti_2As_4O$ ($T_c$ = 21 K) was discovered.[6] The block-layer-design approach was also demonstrated to be effective for other class of layered materials, such as $Eu_3Bi_2S_4F_4$,[7] $LaPbBiS_3O$,[8] $La_5Cu_4As_4O_4Cl_2$,[9] $Pb_5BiFe_3O_{10}Cl_2$,[10] and $Bi_4O_4Cu_{1.7}Se_{2.7}Cl_{0.3}$,[11] etc.

Following the reports of "1144"-type FeSCs, $AkAeFe_4As_4$ ($Ak$ = K, Rb, Cs; $Ae$ = Ca, Sr, Eu),[12-15] we found the first double-$Fe_2As_2$-layer FeSC, $KCa_2Fe_4As_4F_2$ ($T_c$ = 33 K).[16] This finding directly led to the birth of the additional seventeen FeSCs in the "12442"-type family ($T_c$ = 28-37 K).[17-20] The 12442-type compounds can be viewed as the resultant of intergrowth of non-collapsed $ThCr_2Si_2$-type (122-type) $AkFe_2As_2$ and ZrCuSiAs-type (1111-type) CaFeAsF (or, $Re$FeAsO). Our extensive studies[19,20] demonstrate the importance of lattice match, parametrized by the degree of lattice mismatch, $\mu = 2(a_{1111} - a_{122})/(a_{1111} + a_{122})$, where $a_{1111}$ and $a_{122}$ denote the lattice constants of the constituent 1111-type and 122-type compounds. Impressively, all the target 12442 phases can be synthesized with few exceptions, if the criterion $\mu < 2\%$ is satisfied.[20] The empirical criterion remains valid for other types of FeSCs with intergrowth structures.[4] Note that the 12442-type FeSCs are all intrinsically hole doped, which means a substantial charge transfer between the two constituent layers. It is not clear whether such a charge transfer is also essential for the formation of the intergrowth phase. A relevant issue is that whether the parent compound (with the Fe formal valence of 2+) and even the electron-doped FeSC of the 12442 type can be realized.

Recently, we succeeded in preparing a new 1111-type superconductor ThFeAsN ($T_c$ = 30 K) with fluorite-like $Th_2N_2$ spacer layers.[21] We also carried out heavy electron doping up to $x$ =0.6 in $ThFeAsN_{1-x}O_x$, in which a second superconducting region with a maximum $T_c^{max}$ of 17.5 K at $x$ = 0.3 was revealed.[22] The results provide an opportunity to explore an electron-doped 12442-type FeSC. The $a$ axis of $ThFeAsN_{1-x}O_x$ is from 4.037 Å at $x$ = 0 to 3.971 Å at $x$ = 0.6, all satisfying the criterion $\mu < 2\%$ when $BaFe_2As_2$ ($a$ = 3.9625 Å)[23] is taken as the other component for the target 12442-type compound. The large oxygen solubility limit in $ThFeAsN_{1-x}O_x$ allows a wide range of potential charge transfer between the constituent blocks, which may serve as a glue for the formation of the intergrowth phase.[4]

Our trials on the synthesis of $BaTh_2Fe_4As_4(N_{1-x}O_x)_2$ series samples indicated that the oxygen-free target compound, $BaTh_2Fe_4As_4N_2$, could not be prepared although $\mu < 2\%$ is satisfied. For samples of $0.1 \leq x \leq 0.7$, however, the 12442-type phase appears, which strongly suggests the essential role of interlayer charge transfer. The best sample was obtained for $x$ = 0.3. In this Letter, we focus on the synthesis, crystal structure, and the physical properties of $BaTh_2Fe_4As_4(N_{0.7}O_{0.3})_2$. The extended data of the $BaTh_2Fe_4As_4(N_{1-x}O_x)_2$ series samples can be seen in the Supporting Information.

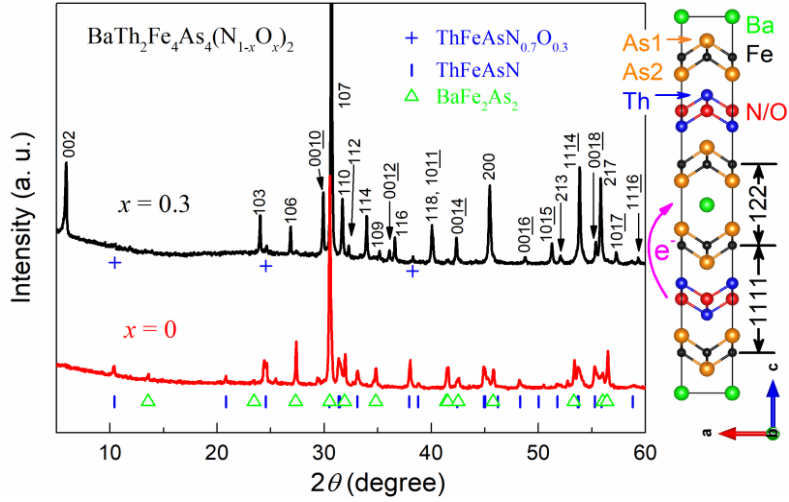

**Figure 1.** X-ray diffraction patterns of the BaTh$_2$Fe$_4$As$_4$(N$_{1-x}$O$_x$)$_2$ samples of with $x = 0$ (bottom) and 0.3 (top). Shown on the right is the expected crystal structure that is stabilized by the charge transfer from the 1111-type block to the 122-type one.

**Results and discussion.** Figure 1 shows the X-ray diffraction (XRD) patterns of BaTh$_2$Fe$_4$As$_4$(N$_{1-x}$O$_x$)$_2$ with $x = 0$ and 0.3. The main phases in the $x = 0$ sample are identified to be ThFeAsN and BaFe$_2$As$_2$. The result is confirmed by the electron-probe micro-analysis (see Table S2 in the Supplementary information). In contrast, most of the XRD reflections of the $x = 0.3$ sample can be well indexed with a 12442-type unit cell, $a \approx (a_{1111} + a_{122})/2$ and $c \approx 2c_{1111} + c_{122}$. Small extra reflections at $2\theta = 24.6°$ and $38.3°$ are probably due to unreacted ThFeAsN$_{0.7}$O$_{0.3}$.[22] The chemical composition in the crystalline grains of BaTh$_2$Fe$_4$As$_4$(N$_{0.7}$O$_{0.3}$)$_2$ is confirmed by the combined study of wavelength-dispersive X-ray spectroscopy and energy-dispersive X-ray spectroscopy (see Tables S1 and S3 in the Supplementary Information). Since the lattice mismatch parameter $\mu$ in BaTh$_2$Fe$_4$As$_4$N$_2$ is 1.8, which satisfies the requirement of lattice match,[4,20] the failure of synthesis of BaTh$_2$Fe$_4$As$_4$N$_2$ conversely suggests the necessity of the charge transfer arising from the oxygen doping. In fact, the 12442-type phase appears in a wide range of the nominal oxygen content ($0.1 \leq x \leq 0.7$, see Fig. S3 in the Supplementary Information). The oxygen doping induces extra electrons to each Fe$_2$As$_2$ layer, equivalent to a charge transfer of $x/2$ from the ThFeAsN$_{1-x}$O$_x$ block to the BaFe$_2$As$_2$ one. This inter-block-layer charge transfer enhances the interlayer Coulomb attractions, which stabilizes the target phase. We conjecture that the stabilization in Bi$_4$O$_4$Cu$_{1.7}$Se$_{2.7}$Cl$_{0.3}$[11] and Pr$_4$Fe$_2$As$_2$Te$_{1-x}$O$_4$[25] may also be due to a similar interlayer charge transfer mechanism.

The crystal structure of BaTh$_2$Fe$_4$As$_4$(N$_{0.7}$O$_{0.3}$)$_2$ was determined by the XRD Rietveld refinement using the crystallographic data of KCa$_2$Fe$_4$As$_4$F$_2$[16] as the initial input. The refinement was successful (Fig. 2) with the reliable factors $R_{wp}$, $R_p$, and $R_B$ all less than 5%, and the result was listed in Table 1. The $a$ axis (3.9886 Å) lies between those of ThFeAsN$_{0.7}$O$_{0.3}$ (3.9955 Å)[22] and BaFe$_2$As$_2$ (3.9625 Å)[23], yet it is 0.24% larger than the average of the latter two $a$ axes. Meanwhile, the $c$ axis (29.853 Å) is close to, but 0.33% lower than, the sum of that of BaFe$_2$As$_2$ (13.0168 Å)[23] and twice of that of ThFeAsN$_{0.7}$O$_{0.3}$ (2 × 8.4683 Å)[22]. The shrinkage in $c$ axis confirms stronger interlayer chemical bonding that stabilizes the intergrowth structure. Similar results were seen in other 12442-type compounds,[16,19,20] nevertheless, the axial ratio ($c/a$ = 7.485) of the present compound is remarkably lower than those (7.9-8.2) of all the previous 12442-type materials. The fundamental origin is that, unlike the previous 12442-type FeSCs that are all hole doped, the present 12442-type BaTh$_2$Fe$_4$As$_4$(N$_{0.7}$O$_{0.3}$)$_2$ is actually electron doped (see the Hall measurement below) in relation with the interlayer charge transfer.

In the inset of Fig. 2 we show two structural parameters of the Fe$_2$As$_2$ layers, the As heights relative to the Fe plane ($h_{As}$) and the As–Fe–As bond angle ($\alpha$). The two parameters were proposed to be associated with $T_c$.[26,27] The highest $T_c$ tends to appear at $h_{As}$ = 1.38 Å or $\alpha$ = 109.5°. Because there are inequivalent sites (As1 and As2) for arsenic atoms in BaTh$_2$Fe$_4$As$_4$(N$_{0.7}$O$_{0.3}$)$_2$, we have two distinct values for $h_{As}$ and $\alpha$. The values of $h_{As1}$ and $h_{As2}$ are 1.302 Å and 1.343 Å, respectively, substantially lower than the optimal one.[26] Meanwhile, $\alpha_1$ and $\alpha_2$ are 113.7° and 112.1°, obviously larger than the ideal value of 109.5°.[27] The deviations from the optimal values may account for the relatively low $T_c$ value (see the result below). Note that the deviations are the opposite in hole-doped 12442-type FeSCs, the latter of which show larger $h_{As}$ and smaller $\alpha$ values (e.g., $h_{As1}$ = 1.405 Å, $h_{As2}$ = 1.436 Å; $\alpha_1$ =108.0°, $\alpha_2$ =106.8° for KCa$_2$Fe$_4$As$_4$F$_2$[16]).

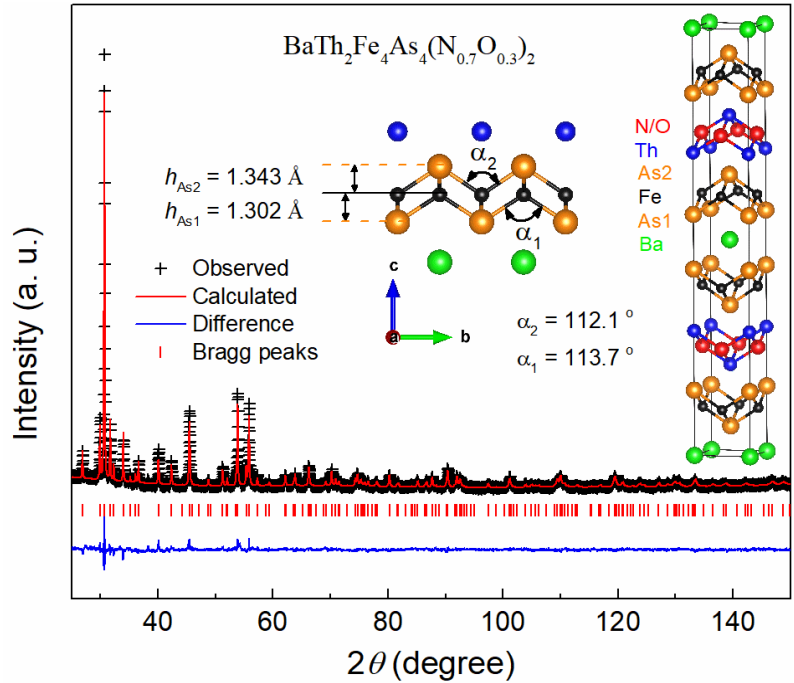

**Figure 2.** Rietveld refinement of the powder X-ray diffraction of BaTh$_2$Fe$_4$As$_4$(N$_{0.7}$O$_{0.3}$)$_2$ from which the crystal structure was determined (shown in the right-hand inset). The middle inset shows the structural parameters of the Fe$_2$As$_2$ layer.

**Table 1.** Crystallographic data for $BaTh_2Fe_4As_4(N_{0.7}O_{0.3})_2$ (Space Group: $I4/mmm$, No. 139) at room temperature. The lattice parameters are $a$ = 3.9886(4) Å and $c$ = 29.853(3) Å.

| atom | site | $x$ | $y$ | $z$ | Occ.[a] | $B_{iso}$[b] |
|------|------|-----|-----|-----|---------|--------------|
| Ba | 2a | 0 | 0 | 0 | 1.0 | 1.7(7) |
| Th | 4e | 0.5 | 0.5 | 0.20839(4) | 1.0 | 0.12(3) |
| Fe | 8g | 0.5 | 0 | 0.10757(13) | 1.0 | 0.70(9) |
| As1 | 4e | 0.5 | 0.5 | 0.06399(11) | 1.0 | 0.19(9) |
| As2 | 4e | 0 | 0 | 0.15261(10) | 1.0 | 0.2 |
| N/O | 4d | 0.5 | 0 | 0.25 | 1.0 | 1.5 |

[a]The occupancy of each atom was fixed to be 1.0 in the Rietveld refinement. [b]The $B$ factors of As2 and N/O were fixed to avoid an unphysical negative value.

Figure 3 shows the temperature dependence of electrical resistivity ($\rho$) for the $BaTh_2Fe_4As_4(N_{0.7}O_{0.3})_2$ polycrystalline sample. The $\rho(T)$ data behave as a conventional metal. Unlike the $\rho(T)$ behavior in hole-doped 12442-type FeSCs,[16-20] here neither a convex curvature at around 150 K nor a linear temperature dependence below 100 K is seen. The low-temperature (40 K < $T$ < 100 K) normal-state resistivity actually satisfies a power relation, $\rho = \rho_0 + AT^n$. The data fitting yields $\rho_0$ = 0.288 mΩ cm, $A$ = 5.23×10$^{-5}$ mΩ cm/K$^{1.87}$, and $n$ = 1.87, coinciding with a Fermi-liquid behavior. The resistivity starts to drop at 30 K, and zero resistivity is achieved at 20 K. The origin of the broad superconducting transition will be discussed later.

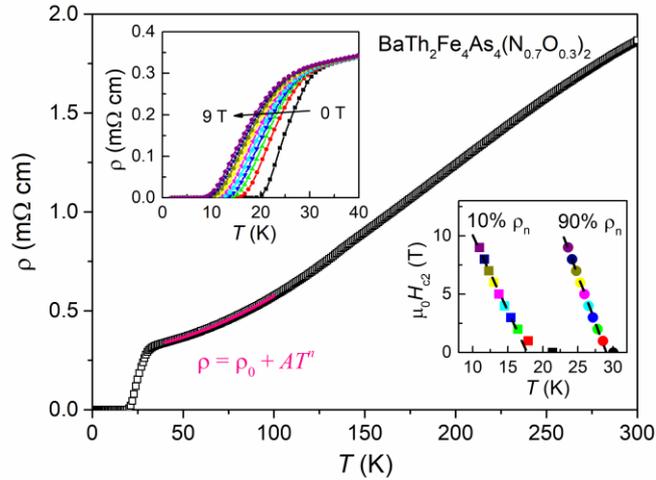

**Figure 3.** Temperature dependence of resistivity for the $BaTh_2Fe_4As_4(N_{0.7}O_{0.3})_2$ polycrystalline sample. The solid pink line represents the data fitting with the formula shown ($n$ = 1.87). The upper-left inset shows the superconducting transitions under external magnetic fields, and the lower-right inset plots the upper critical field ($H_{c2}$) defined in the text.

The superconducting resistive transitions shift to lower temperatures under external magnetic fields. Taken the criterions of resistivity drop to 90% and 10%, the upper critical fields were obtained and plotted in the lower inset of Fig. 3. The $\mu_0 H_{c2}(T)$ data are basically linear. The absolute values of the slopes are fitted to be 1.63 and 1.29 T K$^{-1}$, respectively, both of which are about one order of magnitude smaller than those of the hole-doped 12442-type FeSCs.[16,17,28] The result suggests much longer superconducting coherence length in the electron-doped FeSCs. There are no pronounced resistivity tails under magnetic fields, again different from the case of hole-doped 12442-type FeSCs.[16,17,28] The phenomenon could be understood by the longer coherence length which leads to more three-dimensional superconductivity in BaTh$_2$Fe$_4$As$_4$(N$_{0.7}$O$_{0.3}$)$_2$.

Figure 4(a) shows the superconducting diamagnetic transition in the dc magnetic susceptibility ($\chi$). The onset diamagnetic transition temperature at 30 K is seen in the inset, consistent with the resistivity measurement above. The magnetic shielding volume fraction, measured in ZFC process, is about 100% at temperatures far below $T_c$, suggesting bulk superconductivity. Nevertheless, the value of $4\pi\chi_{ZFC}$ becomes very small at temperatures approaching $T_c$ (e.g., −2.4% at 20 K). The possible explanation is that 30-K superconductivity appears at the crystalline-grain surfaces and/or grain boundaries, and bulk superconductivity seems to emerge at a lower temperature. Indeed, a kink-like anomaly in $\chi_{FC}$ can be observed at 23 K, which could be the sign of bulk superconducting transition.

To demonstrate the bulk superconducting transition, we performed the specific-heat ($C$) measurement. Figure 4(b) shows the $C(T)$ data at zero and 9-T fields. No specific-heat anomaly associated with the superconducting transition can be directly seen in the raw data. Nevertheless, the specific-heat difference, $C_{0T} − C_{9T}$, clearly shows a peak-like feature at 22 K (the negative background could be due to the contribution from a Schottky anomaly under magnetic fields). The peak height ($\Delta C$) is estimated to be 380 mJ K$^{-1}$ mol$^{-1}$, corresponding to $\Delta C/T_c$ = 17 mJ K$^{-2}$ mol-fu$^{-1}$ = 4.3 mJ K$^{-2}$ mol-Fe$^{-1}$. The latter is comparable to the counterpart of LaFeAsO$_{0.9}$F$_{0.1-\delta}$,[29] confirming the bulk superconductivity. One may also see a small anomaly just below 30 K. This seems to be associated with the onset superconducting transitions observed above in $\rho(T)$ and $\chi(T)$.

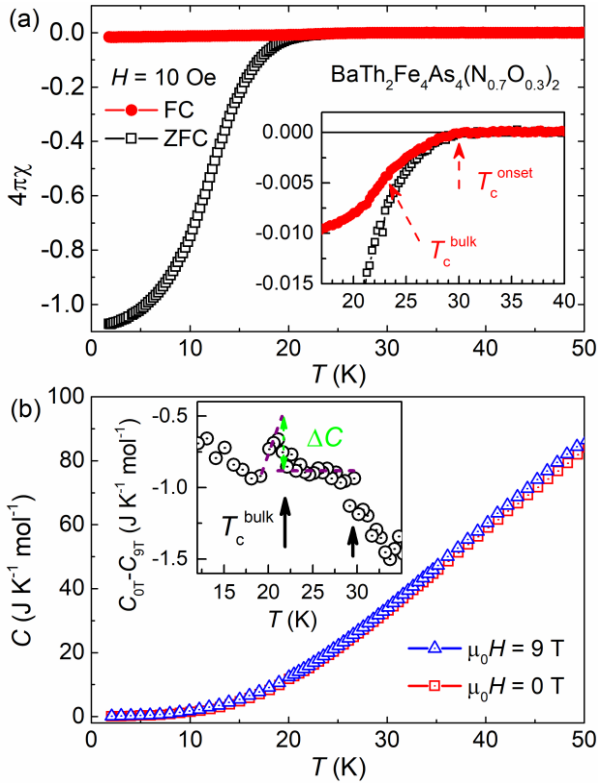

**Figure 4.** Temperature dependence of magnetic susceptibility (a) and specific heat (b) for BaTh$_2$Fe$_4$As$_4$(N$_{0.7}$O$_{0.3}$)$_2$. The top inset shows a close-up for the onset superconducting transition. The bottom inset plots the specific-heat difference, from which a bulk superconducting transition at ~22 K is revealed.

Figure 5 shows the Hall resistance as a function of magnetic field for $BaTh_2Fe_4As_4(N_{0.7}O_{0.3})_2$. The Hall resistance decreases almost linearly for all the data sets measured at different temperatures. The result indicates dominant electron-type conduction, confirming the electron-doping scenario. The Hall coefficient ($R_H$) near $T_c$ is $-2.6\times10^{-9}$ m$^3$ C$^{-1}$, corresponding to a Hall carrier number of 0.14 electrons/Fe within a single-band model (albeit of the multi-band reality). The Hall number coincides with the expected electron concentration of 0.15 electrons/Fe from the nominal composition. The inset of Fig. 5 shows that $R_H$ increases with temperature, probably due to the common multi-band effect in FeSCs. As a comparison, the $R_H$ value of ThFeAsN is $-2.5\times10^{-8}$ m$^3$ C$^{-1}$ at 50 K,[30] an order of magnitude larger, suggesting much fewer electron doping in ThFeAsN than that in $BaTh_2Fe_4As_4(N_{0.7}O_{0.3})_2$.

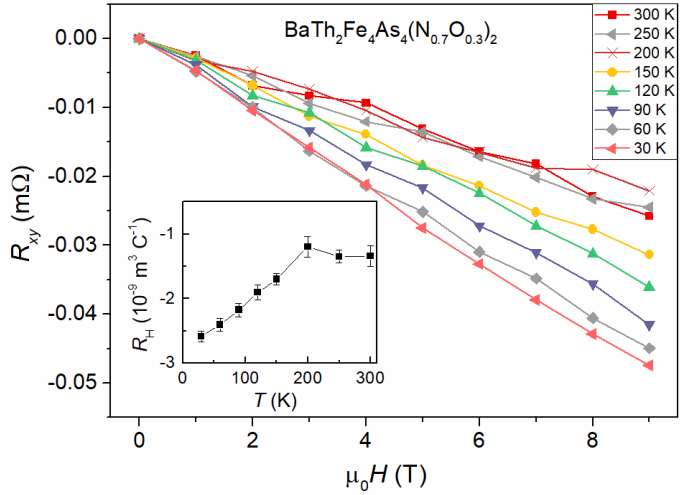

**Figure 5.** Field dependence of Hall resistance of the $BaTh_2Fe_4As_4(N_{0.7}O_{0.3})_2$ sample. The inset plots the Hall coefficient ($R_H$) as a function of temperature.

Because the $Fe_2As_2$ layers in $BaTh_2Fe_4As_4(N_{0.7}O_{0.3})_2$ are electron doped, the emergence of superconductivity seems not surprising.[2,3] However, the single-$Fe_2As_2$-layer material $ThFeAsN_{0.85}O_{0.15}$, which bears the same electron doping level as $BaTh_2Fe_4As_4(N_{0.7}O_{0.3})_2$ does, is not superconducting down to 2 K.[22] In fact, the series samples of $BaTh_2Fe_4As_4(N_{1-x}O_x)_2$ show similar $T_c$ values (see Figs. S5 and S6 in the Supplementary Information). These results suggest that the FeSCs with single and double $Fe_2As_2$ layers can be very different.

**Conclusion.** To summarize, we have discovered the first electron-doped double-$Fe_2As_2$-layer FeSC, $BaTh_2Fe_4As_4(N_{0.7}O_{0.3})_2$, whose $T_c^{onset}$ and $T_c^{bulk}$ are 30 K and 22 K, respectively. The new FeSC shows contrasting differences in the structural details of the $Fe_2As_2$ layers and in the normal-state physical properties, compared with the previous hole-doped 12442-type FeSCs. It also exhibits difference with its single-$Fe_2As_2$-layer counterpart, $ThFeAsN_{0.85}O_{0.15}$. More importantly, we found that this exceptional 12442-type FeSC was stabilized by the inter-block-layer charge transfer. Such interlayer charge transfer serves as an additionally essential role for the formation of an intergrowth structure. This conclusion could be helpful for the exploration of broader layered materials.

**Acknowledgements**   This work was supported by the National Key Research and Development Program of China (Grant No. 2017YFA0303002) and the Fundamental Research Funds for the Central Universities of China (2019FZA3004).

**Author contributions**   Cao GH designed the experiment, discussed the result and wrote the paper. Wang ZC and Wang C also contributed the research idea and discussed the result. Shao YT synthesized the samples, and performed the structural characterizations and physical property measurements with assistance from Li BZ, Wu SQ, Wu JF, Ren Z, Qiu SW, and Rao C.

**Conflict of interest**   The authors declare no competing financial interests.

# Supplementary Information

I. Experimental Details
II. SEM-EDS Results
III. WDS Results
IV. Extended data for BaTh$_2$Fe$_4$As$_4$(N$_{1-x}$O$_x$)$_2$

## I.     Experimental Details

Polycrystalline samples of BaTh$_2$Fe$_4$As$_4$(N$_{1–x}$O$_x$)$_2$ were synthesized by high-temperature solid-state reactions in sealed vacuum. The source materials are ThO$_2$ powders (99.9%), Ba rods (99%), Fe powders (99.998%), and As pieces (99.999%). First, we prepared the intermediate materials Th, Th$_3$N$_4$, Th$_3$As$_4$, FeAs, Fe$_2$As, and BaFe$_2$As$_2$. Fe$_2$As and Th$_3$As$_4$ were prepared by heating their stoichiometric mixtures in sealed quartz ampule to 850 °C, holding for 30 h. BaFe$_2$As$_2$ was prepared by reacting the mixtures of BaAs and Fe$_2$As at 1000 °C for 40 h. Other intermediate materials Th, Th$_3$N$_4$, and FeAs were prepared as reported elsewhere.[21] Second, stoichiometric mixtures of Fe, ThO$_2$, Th$_3$As$_4$, Th$_3$N$_4$, and FeAs, and BaFe$_2$As$_2$ were pressed into pellets, then placed in an alumina tube. The sample-loaded alumina tube was sealed in an evacuated quartz ampule, which was heated to 950 °C holding for 60 hours. The solid-state reactions were repeated with intermediate homogenization by grinding, which could improve the sample's quality.

The Powder X-ray diffraction (XRD) was carried out on a PANalytical X-ray diffractometer with Cu K$\alpha$1 radiation. The crystal structure of BaTh$_2$Fe$_4$As$_4$(N$_{0.7}$O$_{0.3}$)$_2$ was refined by a Rietveld analysis using RIETAN-FP software.[24] The low-angle XRD data ($2\theta <$ 25 °) were not included in the refinement because of the effect of grazing incidence. The nitrogen and oxygen content was fixed to be the nominal one. The final reliable factors $R_{wp}$, $R_p$, and $R_B$ was respectively 4.79%, 3.73%, and 4.46%, and the goodness-of-fit parameter $\chi^2$ was 1.20, indicating the reliability and validity of the refinement. The chemical composition of the specimen with $x = 0$ and 0.3 was determined by electron probe microanalysis using wavelength-dispersive spectroscopy (WDS) and/or energy-dispersive spectroscopy (EDS). The experimental details are given in the Supporting Information.

The electrical transport properties and the specific heat were measured on a Quantum Design Physical Property Measurement System (PPMS Dynacool). A standard four-electrode method was employed for the resistivity measurement. For the Hall measurement, a crosslike four-wire configuration was adopted. The specific-heat data were measured with a thermal relaxation method under zero and 9-T fields, respectively. The dc magnetization was measured on a Quantum Design Magnetic Property Measurement System (MPMS3). The sample was shaped into a regular rod whose demagnetization factor could be easily calculated and corrected. We adopted both the zero-field cooling (ZFC) and the field-cooling (FC) protocols to probe the superconducting diamagnetic transition.

## II. SEM-EDS Results

The BaTh$_2$Fe$_4$As$_4$(N$_{0.7}$O$_{0.3}$)$_2$ polycrystalline sample was examined by the SEM-EDS experiments using a Hitachi S-4800 equipped with an AMETEK© EDAX (Model Octane Plus) spectrometer. Figure S1 shows the typical SEM image, from which plate-like crystalline grains can be seen. The grain size is about 2-5 μm in length and 0.1-0.3 μm in thickness, consistent with the layered crystal structure.

We performed a quantitative element analysis employing the energy-dispersive X-ray spectroscopy (EDS). The electron beam was focused on either a single crystalline grain or an area covering several grains. Figure S2 shows the typical EDS spectrum. The quantitative analysis confirms the stoichiometric ratio except for nitrogen and oxygen (note that there is a systematic positive error of 10% for the As content in our measurement system [1]). In general, EDS gives a large error for light elements such as N and O. We found that the total N/O atomic percentage measured was almost equal to the sum of those of the other elements. This means that the N/O content was much overestimated (note that the N/O adsorptions probably contribute the positive error as well). Therefore, we present the EDS result in Table S1 based on the constrains, $n_{Fe} = 4.00$ and $n_N + n_O = 2.0$, in accordance with the ideal chemical formula. After the correction for the As content [1], the preliminary result of the chemical formula is Ba$_{1.00(8)}$Th$_{1.86(11)}$Fe$_{4.00}$As$_{4.17(10)}$N$_{1.21(6)}$O$_{0.79(6)}$, basically consistent with the nominal composition.

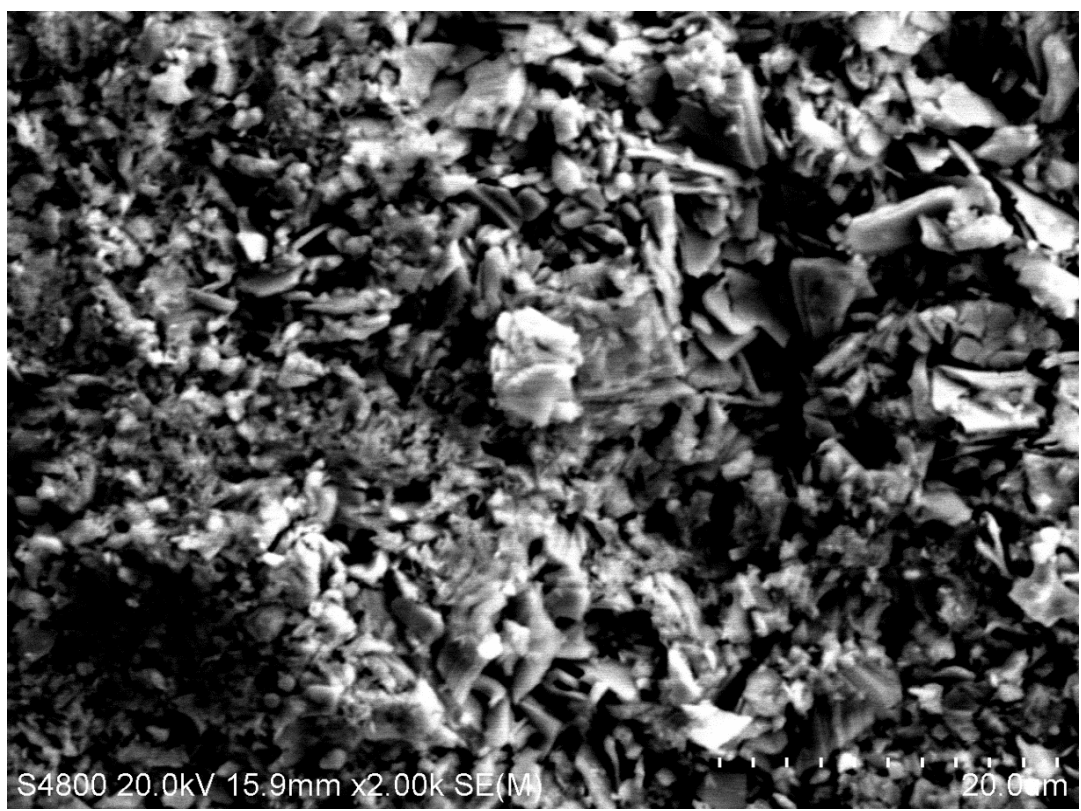

**Figure S1:** A typical SEM image of the BaTh$_2$Fe$_4$As$_4$(N$_{0.7}$O$_{0.3}$)$_2$ polycrystalline sample.

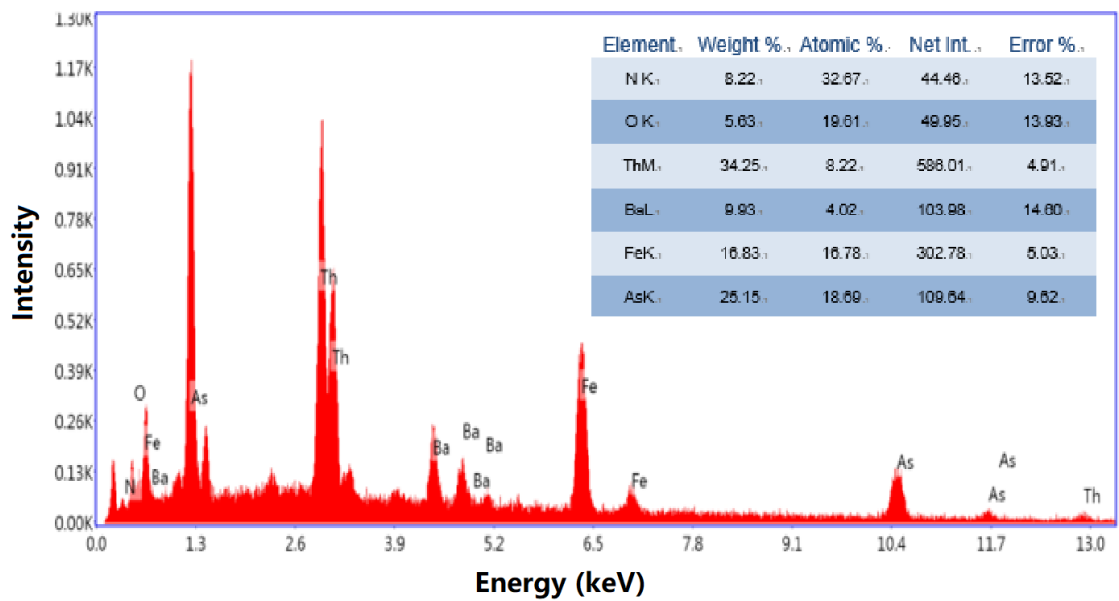

**Figure S2:** Typical energy-dispersive X-ray spectrum of the BaTh$_2$Fe$_4$As$_4$(N$_{0.7}$O$_{0.3}$)$_2$ polycrystalline sample. The inset shows the quantitative element analysis result using eZAF Smart Quant program.

**Table S1:** EDS results for the sample with the nominal composition of BaTh$_2$Fe$_4$As$_4$(N$_{0.7}$O$_{0.3}$)$_2$. The atomic ratios were derived based on the constrains, $n_{Fe} = 4.00$ and $n_N + n_O = 2.0$, in accordance with the ideal chemical formula.

| Elements<br>Exper. # | Ba | Th | Fe | As | N | O |
|---|---|---|---|---|---|---|
| 1 | 1.01 | 1.64 | 4.00 | 4.59 | 1.29 | 0.713 |
| 2 | 0.874 | 2.01 | 4.00 | 4.63 | 1.19 | 0.808 |
| 3 | 1.00 | 1.85 | 4.00 | 4.62 | 1.25 | 0.747 |
| 4 | 0.949 | 1.74 | 4.00 | 4.71 | 1.23 | 0.770 |
| 5 | 1.15 | 1.92 | 4.00 | 4.68 | 1.10 | 0.897 |
| 6 | 1.06 | 1.82 | 4.00 | 4.47 | 1.12 | 0.881 |
| 7 | 0.916 | 1.87 | 4.00 | 4.72 | 1.22 | 0.78 |
| 8 | 0.999 | 1.90 | 4.00 | 4.43 | 1.23 | 0.766 |
| 9 | 0.958 | 1.96 | 4.00 | 4.46 | 1.24 | 0.764 |
| Mean value | 0.991 | 1.86 | 4.00 | 4.59 | 1.21 | 0.792 |
| Standard deviation | 0.0765 | 0.107 | -- | 0.105 | 0.0581 | 0.0573 |

### III. WDS Result

To determine the N/O content more precisely and accurately, we also employed the technique of wavelength-dispersive X-ray spectroscopy (WDS) with an EPMA 1720H (Shimadzu) electron microprobe. The samples were polished with ultrafine silicon-carbide paper before the measurements. The accelerating voltage was 15 kV, the beam current was 15 nA, and the beam size was no larger than 5 µm (in diameter). Natural minerals were used as the standards. A program based on ZAF4 procedure was applied for data correction.

Here we report the WDS results for two samples with the nominal composition of

BaTh$_2$Fe$_4$As$_4$N$_2$ and BaTh$_2$Fe$_4$As$_4$(N$_{0.7}$O$_{0.3}$)$_2$, respectively. As we know from the main article, the "BaTh$_2$Fe$_4$As$_4$N$_2$" sample actually consists of two phases, BaFe$_2$As$_2$ and ThFeAsN. Indeed, the WDS result shown in Table S2 indicates two different compositions (in different image contrast), pointing to the phases of BaFe$_2$As$_2$ and ThFeAsN, respectively. Note that we always detected some oxygen, even in the region of BaFe$_2$As$_2$. For the ThFeAsN region, the oxygen content is about 20% of the nitrogen content. This means that the N/O-site occupancy would be about 120%, which is unrealistic. For ThFeAsN, as a matter of fact, the nitrogen content is determined to be nearly 1.0 (either 0.97 with 2.7% nitrogen vacancy or 0.93 with 7% oxygen substitution) by the neutron diffraction study [2]. Therefore, the extra oxygen detected is probably due to the oxygen adsorption. The oxygen adsorption was unavoidable when one handles the samples in air.

**Table S2** WDS results for the sample with the nominal composition of BaTh$_2$Fe$_4$As$_4$N$_2$. The atomic ratios of the two kinds of regions, which are respectively dominated from BaFe$_2$As$_2$ and ThFeAsN, were derived based on the constrains, $n_{Fe}$ = 2.00 and $n_{Fe}$ = 1.00. The oxygen is not taken into the consideration because of the possible oxygen adsorption.

| Type-I region | | | | | | |
|---|---|---|---|---|---|---|
| Elements / Exper. # | Ba | Th | Fe | As | N | Note |
| 1 | 0.8995 | 0.01863 | 2.00 | 1.895 | 0.1142 | BaFe$_2$As$_2$ phase dominated |
| 2 | 0.9115 | 0.01102 | 2.00 | 1.678 | 0.1292 | |
| Mean value | 0.9055 | 0.01483 | 2.00 | 1.787 | 0.1217 | |
| Standard deviation | 0.006 | 0.00380 | -- | 0.109 | 0.00750 | |
| Chemical formula | Ba$_{0.91(1)}$Th$_{0.015(4)}$Fe$_{2.00}$As$_{1.79(11)}$N$_{0.122(8)}$ | | | | | |
| Type-II region | | | | | | |
| Elements / Exper. # | Ba | Th | Fe | As | N | Note |
| 3 | 0.03010 | 0.9973 | 1.00 | 1.002 | 0.9013 | ThFeAsN phase dominated |
| 4 | 0.01667 | 1.009 | 1.00 | 0.9105 | 1.090 | |
| 5 | 0.01674 | 0.9963 | 1.00 | 0.9530 | 1.118 | |
| Mean value | 0.02117 | 1.001 | 1.00 | 0.9552 | 1.036 | |
| Standard deviation | 0.00632 | 0.00577 | -- | 0.0374 | 0.0962 | |
| Chemical formula | Ba$_{0.021(1)}$Th$_{1.00(1)}$Fe$_{1.00}$As$_{0.96(4)}$N$_{1.04(10)}$ | | | | | |

Table S3 presents the WDS result of the BaTh$_2$Fe$_4$As$_4$(N$_{0.7}$O$_{0.3}$)$_2$ sample. The derived chemical formula is Ba$_{1.03(1)}$Th$_{2.05(8)}$Fe$_{4.00}$As$_{3.95(42)}$N$_{1.42(32)}$O$_{1.38(19)}$ on the basis of $n_{Fe}$ = 4.00. Note that the sum of the N/O content is 2.80(51), which is remarkably larger than the ideal value of 2.00. The above EDS result for BaTh$_2$Fe$_4$As$_4$(N$_{0.7}$O$_{0.3}$)$_2$ and the WDS data for "BaTh$_2$Fe$_4$As$_4$N$_2$" strongly suggest the oxygen/nitrogen adsorption. If the nitrogen adsorption is neglected (owing to the inertness of nitrogen), one concludes that the chemical formula measured is Ba$_{1.03(1)}$Th$_{2.05(8)}$Fe$_{4.00}$As$_{3.95(42)}$N$_{1.42(32)}$O$_{0.58(19)}$, fully consistent with the nominal one within the measurement uncertainties.

**Table S3** WDS results for the sample with the nominal composition of BaTh$_2$Fe$_4$As$_4$(N$_{0.7}$O$_{0.3}$)$_2$. The final chemical formula was derived based on $n_{Fe}$ = 4.00. Note that sum of the N/O content is obviously larger than the ideal value because of the primary oxygen adsorption.

| Elements / Exper. # | Ba | Th | Fe | As | N | O |
|---|---|---|---|---|---|---|
| 1 | 1.032 | 2.026 | 4.00 | 3.683 | 1.568 | 1.483 |
| 2 | 0.9110 | 2.041 | 4.00 | 3.520 | 1.465 | 1.572 |
| 3 | 1.211 | 2.056 | 4.00 | 3.866 | 1.456 | 1.478 |
| 4 | 1.014 | 2.132 | 4.00 | 4.173 | 1.105 | 0.995 |
| 5 | 0.9210 | 1.965 | 4.00 | 3.875 | 1.129 | 1.154 |
| 6 | 1.055 | 2.124 | 4.00 | 4.184 | 1.174 | 1.449 |
| 7 | 1.050 | 2.080 | 4.00 | 3.295 | 2.109 | 1.550 |
| 8 | 0.9958 | 2.021 | 4.00 | 3.839 | 1.825 | 1.610 |
| 9 | 1.088 | 1.909 | 4.00 | 4.880 | 1.098 | 1.278 |
| 10 | 1.004 | 2.182 | 4.00 | 4.210 | 1.272 | 1.259 |
| Mean value | 1.028 | 2.054 | 4.00 | 3.953 | 1.420 | 1.383 |
| Standard deviation | 0.0808 | 0.0769 | -- | 0.418 | 0.322 | 0.192 |
| Chemical formula | Ba$_{1.03(1)}$Th$_{2.05(8)}$Fe$_{4.00}$As$_{3.95(42)}$N$_{1.42(32)}$O$_{1.38(19)}$ | | | | | |

## IV. Extended data for BaTh$_2$Fe$_4$As$_4$(N$_{1-x}$O$_x$)$_2$

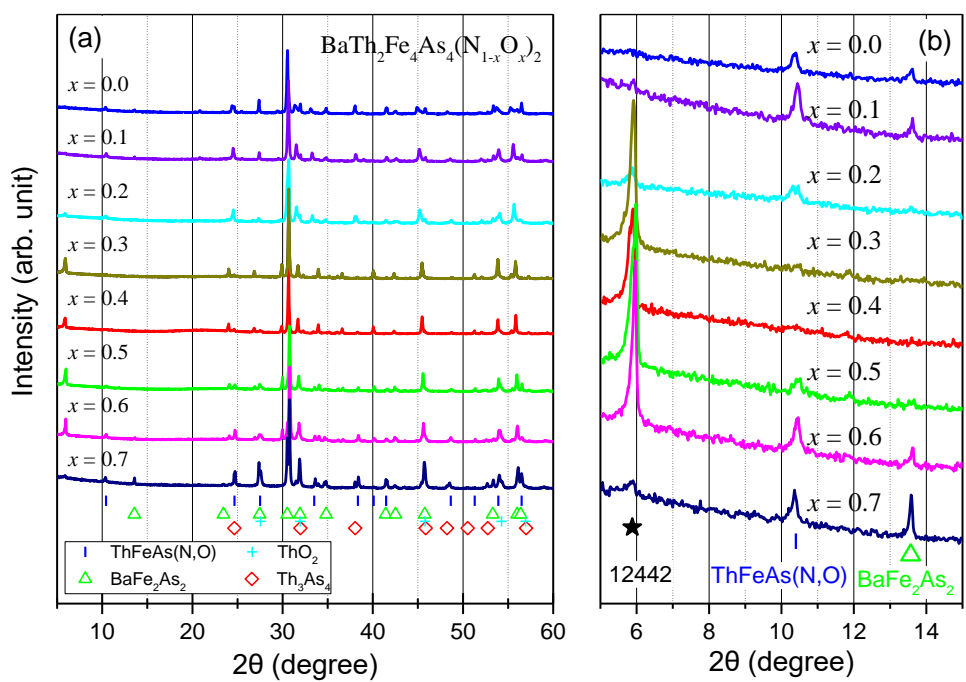

**Figure S3:** (a) Powder X-ray diffraction patterns for the series samples of BaTh$_2$Fe$_4$As$_4$(N$_{1-x}$O$_x$)$_2$. Panel (b) shows a close-up in the low-angle region where the characteristic reflections of the 12442-, 1111-, and 122-type phases can be clearly seen.

Figure S3 shows the powder XRD patterns of the BaTh$_2$Fe$_4$As$_4$(N$_{1-x}$O$_x$)$_2$ series samples. The 12442-type phase forms for $0.1 \leq x \leq 0.7$. Nevertheless, only small fraction of the

12442-type phase was obtained for $x$ = 0.1 and 0.7. We conjecture that the amount of inter-block-layer charge transfer is not high enough in the case of $x$ = 0.1. On the other hand, the 1111-type block layer itself becomes instable for high $x$ values, thus the best samples were obtained for the intermediate values of $x$ = 0.3 and 0.4.

The lattice parameters of the obtained 12442-type phase were calculated based on the XRD patterns, which are plotted in Fig. S4. One sees that, except for $x$ = 0.2 where the 12442-type compound is not the main phase, $a$ and $c$ axes both decrease with the oxygen doping, primarily due to smaller size of $O^{2-}$ compared with $N^{3-}$. Note that the $c$ axis deviates from the expected value more and more with the oxygen doping, further reflecting stronger inter-block-layer Coulomb attractions at higher doping levels.

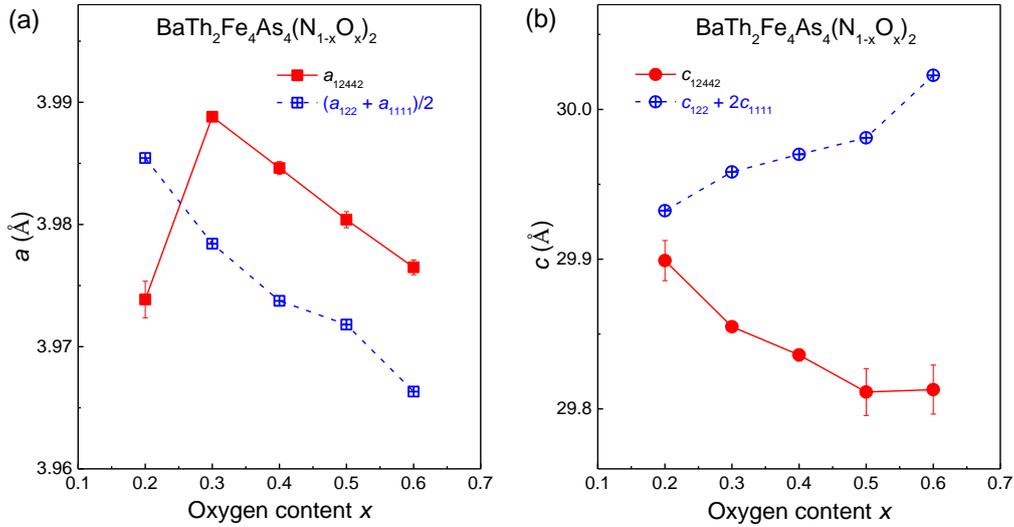

**Figure S4:** Lattice parameters of $BaTh_2Fe_4As_4(N_{1-x}O_x)_2$ as a function of nominal oxygen doping level. The symbols in blue with dashed lines denote the average values of their constituent 122- and 1111-type unit cells.

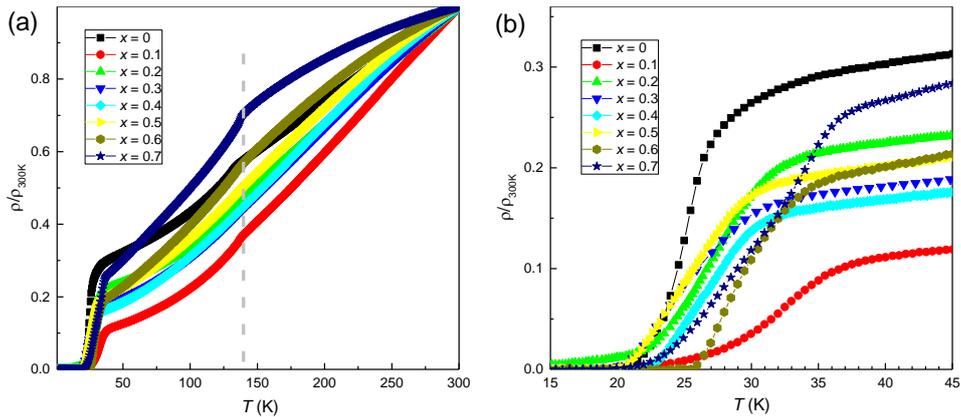

**Figure S5:** Temperature dependence of resistivity (normalized to the value at 300 K) for the $BaTh_2Fe_4As_4(N_{1-x}O_x)_2$ samples. The right panel shows a close-up, highlighting the superconducting transitions.

The resistivity measurement results for the series samples of $BaTh_2Fe_4As_4(N_{1-x}O_x)_2$ are plotted in Fig. S5. All the samples show a superconducting transition with a relatively broad transition width. Note that the superconducting transition for $x = 0$ is due to the superconductivity in the 1111-type phase. The resistivity kinks at about 140 K, marked with a dashed line, is observable for $x$ = 0, 0.1, 0.6, and 0.7, which are attributed to the spin-density-wave transition in $BaFe_2As_2$.

Figure S6 shows the temperature dependence of magnetic susceptibility for the $BaTh_2Fe_4As_4(N_{1-x}O_x)_2$ samples. Bulk superconductivity in the 12442-type phase is confirmed for $x$ = 0.2-0.6, as is clearly seen in the right panel (right axis). Interestingly, we found that the onset transition temperatures, which possibly reflects the interface/surface superconductivity, show a "U"-shape doping dependence. The resistive offset transition temperature may represent the bulk transition temperature, which tends to increase with the electron doping.

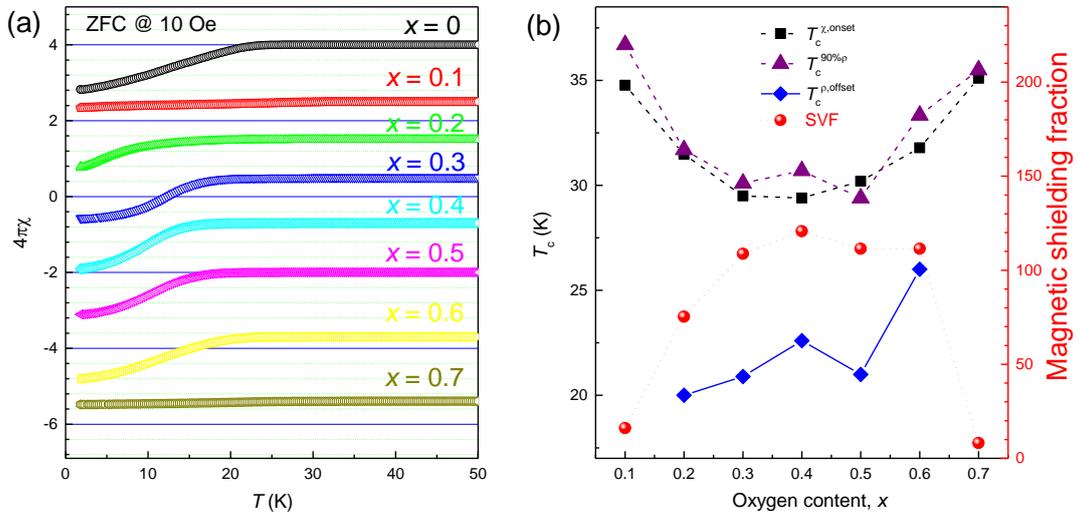

**Figure S6:** (a) Temperature dependence of dc magnetic susceptibility measured in zero-field-cooling mode for the $BaTh_2Fe_4As_4(N_{1-x}O_x)_2$ samples. (b) The derived superconducting (onset) transition temperature as a function of the nominal oxygen doping. The red symbols plotted using the right axis denote the superconducting volume fraction (SVF) defined by the diamagnetic signals at 2 K. The superconducting transition temperatures, $T_c^{90\%\rho}$ and $T_c^{offset}$, from the resistivity measurement are also plotted for comparison.